\documentclass[fleqn,twoside]{article}
\usepackage{espcrc2}

\usepackage{graphicx}
\usepackage[figuresright]{rotating}

\newcommand{\AmS}{{\protect\the\textfont2
  A\kern-.1667em\lower.5ex\hbox{M}\kern-.125emS}}

\title{Weak Phases and CP Violation}

\author{Michael Gronau\,\address{\,Physics Department, Technion -- Israel 
Institute of Technology, \\ 
        32000 Haifa, Israel}}%
       
\begin{document}

\def \app{D_{\pi \pi}}
\def \b{{\cal B}}
\def \bbpp{\overline{{\cal B}}_{\pi \pi}}
\def \bea{\begin{eqnarray}}
\def \beq{\begin{equation}}
\def \bg{\bar \Gamma}
\def \bl{\bar \lambda}
\def \bo{B^0}
\def \ko{K^0}
\def \ob{\overline{B}^0}
\def \lesssim{\stackrel{<}{\sim}}
\def \largesim{\stackrel{>}{\sim}}
\def \bpb{\stackrel{(-)}{B^0}}
\def \cn{Collaboration}
\def \cpp{C_{+-}}
\def \eea{\end{eqnarray}}
\def \eeq{\end{equation}}
\def \ite{{\it et al.}}
\def \kpb{\stackrel{(-)}{K^0}}
\def \lpp{\lambda_{\pi \pi}}
\def \ob{\overline{B}^0}
\def \ok{\overline{K}^0}
\def \rpp{R_{\pi \pi}}
\def \rt{r_{\tau}}
\def \s{\sqrt{2}}
\def \half{\frac{1}{2}}
\def \3half{\frac{3}{2}}
\def \spp{S_{+-}}

\begin{abstract}
Time-dependent CP asymmetries in $B^0$ decays from interference of $\bo$-$\ob$ 
mixing and $b\to c\bar cs,~u\bar ud,~sq\bar q$, and direct asymmetries from 
interference of $b\to c\bar us$ and $b\to u\bar cs$ are studied, improving errors in 
the weak phases $\beta$ and $\gamma$ to $\pm 1^\circ$ and $\pm 7^\circ$, respectively. 
Two kinds of tests for New Physics are discussed: 1. Deviations of asymmetries in 
$b\to sq\bar q$ decays from $\sin 2\beta\sin\Delta mt$. 2.  Violation of
a sum rule for direct asymmetries in $B\to K\pi$. 

\vspace{1pc}
\end{abstract}

\maketitle

\section{INTRODUCTION}

At the end of the second millennium, thirty five years after the discovery in 1964 of CP 
violation in $K\to \pi^+\pi^-$~\cite{Cronin}, theoretical interpretations within the 
Kobayashi-Maskawa framework~\cite{KM} of CP 
nonconservation in $K$ decays involved large hadronic uncertainties.  
This situation has changed dramatically through progress made in the past five 
years by the BaBar  and Belle detectors operating at SLAC and KEK.
Theoretical ideas proposed between twenty five and fifteen years ago and developed 
subsequently to measure the 
weak phases $\beta,~\alpha$ and $\gamma$ through CP asymmetries in 
$B^0\to J/\psi K_S$~\cite{BS}, $B^0\to\pi^+\pi^-,~\rho^+\rho^-$~\cite{GL} and 
$B^+\to DK^+$~\cite{GW} were applied experimentally, thereby improving greatly our 
confidence in the Kobayashi-Maskawa mechanism of CP violation. 
The purpose of this presentation is to describe this remarkable progress, applying 
simple equations (instead of $\chi^2$ fits) to most recent data in 
order to obtain the current values of $\beta, \alpha$ and $\gamma$.

Two major targets of high statistics experiments studying $B$ and $B_s$ decays in $e^+e^-$
and hadron colliders are (1) achieving great precision in Cabibbo-Kobayashi-Maskawa (CKM)
parameters, and (2) identifying potential inconsistencies by overconstraining these parameters. 
For instance, the phase $2\beta$ measured in time-dependent CP asymmetries of $B^0$ 
decays via $b \to c \bar cs$ may be tested also in $b \to sq\bar q~(q=u,d,s)$ penguin-dominated decays~\cite{PenLP,PenMG} which are susceptible to effects of physics beyond the Standard Model~\cite{GroWo}. 

Section 2 reviews the current status of precision determinations of the weak phases 
$\beta,~\alpha$ and $\gamma$, while Section 3 compares measurements of $\sin 2\beta$
in $b\to c\bar cs$ and in penguin-dominated decays. A way of identifying New
Physics in the latter decays through direct CP asymmetries in $B\to K\pi$ is discussed in 
Section 4, and Section 5 concludes with a few remarks about future prospects. 

\section{PRECISION TESTS FOR $\beta,~\alpha,~\gamma$}

Semileptonic $B$ decays, $B^0$-$\bar B^0$ mixing, $B_s$-$\bar B_s$ 
mixing, and $\epsilon_K$ constrain indirectly the three angles of the unitarity 
triangle in an overall fit combining theoretical 
and experimental errors~\cite{CKMfit},
\beq
\beta = (23\pm 4)^\circ,~~\alpha = (99 \pm 13)^\circ,~~\gamma =(58 \pm 13)^\circ.
\eeq
We quote symmetric $1\sigma$ errors using CKMfitter Group pre-summer 2005. 
A crucial question confronting certain measurements of CP asymmetries  in $B$ 
decays is do they agree with these values and can they reduce the above errors?  

\subsection{The phase $\beta$}

\medskip
The classical way of determining $\sin 2\beta$ in time-dependent CP asymmetries~\cite{BS} 
is based on interference between $\bo$-$\bar{B^0}$ mixing and a $b\to c\bar cs$ decay amplitude
which carries a single weak phase at a very high precision~\cite{PenMG,Mannel}. The world 
averaged value before Summer 2005~\cite{HFAG}, $\sin 2\beta = 0.726 \pm 0.037$,  
corrected by a recent Belle measurement~\cite{Belle}, $\sin 2\beta = 0.652 \pm 0.039 
\pm 0.020$, becomes 
\beq
\sin 2\beta = 0.687 \pm 0.032~~.
\eeq

A twofold ambiguity in $\beta$ ($\beta \to \pi/2 -\beta$) in the range 
$0 < \beta < \pi/2$ may be resolved by measuring the sign of $\cos 2\beta$. 
A transversity analysis of $B^0 \to J/\psi K^*$~\cite{Ba_psiK*} implying 
$\cos 2\beta = 2.72^{+0.50}_{-0.79}\pm 0.27$ excludes at $86\%$ 
confidence level a negative value of $\cos 2\beta$. A recent time-dependent 
Dalitz analysis of $D^0\to K_S\pi^+\pi^-$ in $B^0$ decays to $D^0$ and 
a light neutral meson~\cite{Be_Dh0} improves the confidence level to 97$\%$. 
Thus, we conclude $\beta = (21.6 \pm 1.3)^\circ$, in agreement with (1), 
and implying an overall average, $\beta=(21.7\pm 1.2)^\circ$.
[$\beta=(23.2\pm 0.9)^\circ$ was obtained this summer by CKMfitter~\cite{CKMfit}, 
using a constraint from $|V_{ub}|/|V_{cb}|$ in which a very small error $(<5\%)$ was 
assumed.]

\subsection{The phase $\alpha$ in $B\to\pi\pi, \rho\rho, \rho\pi$}

\medskip\noindent
{\it 2.2.1 $B\to\pi\pi$}

The amplitude for $B^0\to\pi^+\pi^-$ contains two terms~\cite{PenLP,PenMG}, 
conventionally denoted ``tree" ($T$) and ``penguin" ($P$) amplitudes, 
involving a weak phase $\gamma$ and a strong phase $\delta$: 
\beq
A(\bo \to \pi^+ \pi^-) =  |T| e^{i \gamma} + |P| e^{i \delta}.
\eeq
Time-dependent decay rates, for an initial $B^0$ or a $\ob$, are
given by~\cite{PenMG}
\bea\label{SC}
&&\Gamma(B^0(t)/\ob(t)\to\pi^+\pi^-) \propto 
e^{-\Gamma t}\Gamma_{\pi\pi}\times\nonumber\\
&&
\left [ 1\pm \cpp\cos\Delta m t \mp \spp\sin\Delta m t\right ],
\eea 
\bea
\spp&= &\frac{2 {\rm Im}(\lpp)}{1 + |\lpp| ^2},~~~
\cpp = \frac{1 - |\lpp|^2}{1 + |\lpp|^2},\\
\lpp & \equiv &  e^{-2i \beta} \frac{A(\ob \to \pi^+ \pi^-)}
{A(B^0 \to \pi^+ \pi^-)}.
\eea

The measurables, $\Gamma_{\pi\pi}$, $\spp$ and $\cpp$ are 
insufficient for determining $|T|, |P|, \delta$ and $\gamma$. One uses 
additional information obtained from an isospin amplitude triangle for $B$ decays,
\beq\label{isotr}
A(\pi^+\pi^-)/\s + A(\pi^0\pi^0)-A(\pi^+\pi^0)=0,
\eeq
and a similar one for $\bar B$. Defining $\sin 2\alpha_{\rm eff} \equiv 
S_{+-}/(1 - C^2_{+-})^{1/2}$, the difference $\theta \equiv \alpha_{\rm eff}-\alpha$ 
is determined up to a sign ambiguity by constructing the two isospin triangles~\cite{GL}. 
The strongest bound on $|\theta|$ in terms of CP-averaged rates and a direct CP asymmetry 
in $B^0\to\pi^+\pi^-$ is~\cite{GQ}
\beq
\cos 2\theta \ge \frac{\left( {1\over 2}\Gamma_{+-} + \Gamma_{+0} - \Gamma_{00} \right)^2 -
 \Gamma_{+-}\Gamma_{+0}}{\Gamma_{+-} \Gamma_{+0} \sqrt{1-C^2_{+-}}}~.
\eeq
This bound is improved by measuring $C_{00}\equiv -$ $A_{CP}(\pi^0\pi^0)$, the direct 
asymmetry in $B^0\to\pi^0\pi^0$.
 
Current measurements~\cite{HFAG}, $\b_{+-} = 
5.0 \pm 0.4$~\footnote{Branching ratios will be quoted in units of $10^{-6}$.}, 
$\b_{+0} = 5.5 \pm 0.6,~\b_{00} = 1.45 \pm 0.29$,
$S_{+-}=-0.50$ $\pm 0.12,~C_{+-}=-0.37\pm 0.10,~C_{00}=-0.28\pm 0.29$, imply 
$\alpha_{\rm eff} = (106\pm 5)^\circ,~|\theta| < 36^\circ$. A second solution at 
$\alpha_{\rm eff}=164^\circ$ is excluded by (1). A positive sign, $\theta\ge 0$, is 
determined~\cite{GLW} by two properties, $|P/T| \le1,~|\delta| \le \pi/2$, which
are confirmed experimentally. This leads to a solution $\alpha = (88\pm 18)^\circ$ 
using isospin symmetry alone. 
[A value $\alpha = (99 \pm 18)^\circ$~\cite{GRalpha}, obtained by relating 
$B\to\pi\pi$ and $B\to K\pi$, does not  contain SU(3) breaking and will not be 
used.] We stress that the condition $\theta \ge 0$ should also 
be applied in $\chi^2$ likelihood fits for $B\to\pi\pi$~\cite{CKMfit}.  

\medskip\noindent
{\it 2.2.2 $B\to\rho\rho$}

Angular analyses of the pions in $\rho$ decays have shown that 
$B^0\to \rho^+\rho^-$ is dominated by longitudinal polarization, $f^{+-}_L = 0.978 \pm 
0.014^{+0.021}_{-0.029}$~\cite{Barr}, 
$0.951^{+0.033}_{-0.039}~^{+0.029}_{-0.031}$~\cite{Berr}. This simplifies the study of 
CP asymmetries in these decays to the level of studying asymmetries in $B^0\to\pi^+\pi^-$.
As long as a nonzero asymmetry $C_{00}$ has not been measured, an advantage of $B\to \rho\rho$ over $B\to \pi\pi$ is  $\b(\rho^0\rho^0)/\b(\rho^+\rho^-)
<\b(\pi^0\pi^0)/\b(\pi^+\pi^-)$, implying a stronger bound on $|\theta|$ in $B\to\rho\rho$.

Using~\cite{HFAG} $\b(\rho^+\rho^-)=26.2^{+3.6}_{-3.7},~\b(\rho^+\rho^0) = 26.4^{+6.1}_{-6.4}$  and~\cite{Bar0r0} $\b(\rho^0\rho^0) < 1.1$, and taking averages for CP
asymmetries from~\cite{Barr} and~\cite{Berr}, $S_L = -0.21 \pm 0.23,~C_L = -0.02 \pm 0.17$,
one finds $\alpha_{\rm eff} = (96 \pm 6)^\circ,~|\theta| < 11^\circ (1\sigma)$, implying
$\alpha = (96 \pm 13)^\circ$. This is currently the most precise single determination of $\alpha$.

\medskip\noindent
{\it 2.2.3 $B\to\rho\pi$}

A time-dependent Dalitz analysis of $B^0\to \pi^+\pi^-\pi^0$ involving interference of 
$B^0\to \rho^\pm\pi^\mp, \rho^0\pi^0$ provides twenty seven mutually dependent 
measurables depending on twelve parameters including $\alpha$~\cite{QS}.
Limited statistics leads to a large statistical error, and potential contributions from 
S-wave $\pi\pi$ resonance states and excited $\rho$ meson states lead to model-dependent 
uncertainties. An analysis  by the Babar collaboration obtained~\cite{Brp} $\alpha= 
(113^{+27}_{-17}\pm 6)^\circ$.

Alternatively, one may apply flavor SU(3) symmetry to quasi two-body 
$B^0\to\rho^{\pm}\pi^{\mp}$, using time-dependent decay rates~\cite{MGPLB},
\bea\label{Gammat}
&\Gamma(B^0(t) \to \rho^\pm\pi^\mp)\propto e^{-\Gamma t} \times [1+\nonumber\\
&(C \pm \Delta C)\cos\Delta mt-(S \pm \Delta S)\sin\Delta mt ]. 
\eea
As in $B\to\pi^+\pi^-$, one defines $\alpha_{\rm eff}$ which 
equals $\alpha$ in the limit of vanishing penguin amplitudes~\cite{GZrp}, 
\bea
4\alpha_{\rm eff} \equiv  \arcsin\left [(S + \Delta S)/\sqrt{1- (C + \Delta C)^2}\right ]
\nonumber\\
 + \arcsin\left [(S - \Delta S)/\sqrt{1- (C - \Delta C)^2}\right ].
 \label{alphaeff}
\eea
In the SU(3) limit, the difference $|\alpha_{\rm eff} - \alpha|$ is bounded by ratios of  
decay rates for $B\to K^*\pi$ and $B\to K\rho$
and decay rates for $B\to \rho^\pm\pi^\mp$. The bound~\cite{GZrp}, 
$|\alpha_{\rm eff}-\alpha|<13^\circ$, can be reduced by a factor two under very 
mild assumptions about (small) ratios of penguin and tree amplitudes and 
about a strong phase difference~\cite{GLW}.

Taking averages from Refs.~\cite{Barp} and \cite{Berp},
\bea
C = 0.30 \pm 0.13,~~\Delta C &=& 0.33 \pm 0.13,\nonumber\\
S = -0.04 \pm 0.17,~~\Delta S &=& -0.07 \pm 0.18,
\eea
one find $\alpha_{\rm eff} = (92 \pm 8)^\circ$ and therefore 
$\alpha=(92 \pm 8 \pm 8)^\circ$. The second error 
follows from the SU(3) bound in which a 30$\%$ uncertainty
is included.  To be conservative, we add the experimental and theoretical 
errors linearly, $\alpha = (92 \pm 16)^\circ$.

\medskip\noindent
{\it 2.2.4 Averaged $\alpha$}

Combining the values of $\alpha$ from $B\to\pi\pi$ and 
$B\to\rho\rho$, one finds an average $\alpha = (93 \pm 11)^\circ$. 
The average becomes $(97 \pm 8)^\circ$ when including 
the two values of $\alpha$ obtained from $B\to \rho\pi$.
This direct determination agrees with Eq.~(1) representing all other CKM 
constraints, and is already more precise than this indirect value. 
Combining these two values we find an overall average 
\beq
\alpha = (98 \pm 7)^\circ.
\eeq
In comparison, two global fits combining all constraints and using different 
methods for error estimates obtain~\cite{CKMfit} $\alpha = (98.1^{+6.3}_{-7.0})$
and~\cite{Ufit}, $(97.9 \pm 6.0)^\circ$.

\medskip\noindent
{\it 2.2.5 Isospin breaking corrections in $\alpha$}

The overall determination $\alpha = (98\pm 7)^\circ$, equivalently  
$\gamma = (60 \pm 7)^\circ$ when using $\beta = (22\pm 1)^\circ$,  
relies in part on isospin symmetry. At this precision one must consider 
isospin breaking corrections caused by the charge and mass differences 
of $u$ and $d$ quarks. Here we will summarize briefly the results of 
a recent study of isospin violating effects in $\alpha$~\cite{GZI} updating an earlier 
analysis~\cite{Gardner}. 

Effects due to the different charges of the $u$ and $d$ quarks have been 
calculated model-independently and process-independently~\cite{BFGPY}  
by noting that the $\Delta I=3/2$  electroweak penguin (EWP) operator in the effective 
Hamiltonian for $b \to dq\bar q$ is proportional to the $\Delta I=3/2$ current-current 
operator (contributions of EWP operators with small 
Wilson coefficients $c_7$ and $c_8$ are neglected). The calculated EWP correction to 
$\alpha$ in $B\to\pi\pi$ and $B\to \rho\rho$ is negative, $\Delta^{EWP}_{\alpha}=
[3(c_9 + c_{10})/2(c_1 + c_2)][\sin(\beta + \alpha)\sin\alpha/\sin\beta]=-(1.7 \pm 
0.3)^\circ$, and should be included in the extracted value using $\alpha = 
\alpha_{\rm eff} - \theta + \delta^{EWP}_{\alpha}$.

Effects caused by $\pi^0$ mixing with $\eta$ and $\eta'$ are
parametrized  in terms of mixing parameters $\epsilon$, $\epsilon'$ of order 
0.01~\cite{Kroll}.
In an SU(3) symmetry expansion their leading effect on the isospin triangle 
(\ref{isotr}) is multiltiplying $A(B^+\to\pi^+\pi^0)$ 
by $1-e_0$, where $e_0 = \epsilon\sqrt{2/3} + \epsilon'\sqrt{1/3} =
0.016 \pm 0.003$. Using measured branching ratios for $B^+\to\pi^+\eta$
and $B^+\to \pi^+\eta'$, one finds a stringent upper limit on the effect of 
$\pi^0$-$\eta$-$\eta'$ mixing on 
$\alpha$~\cite{GZI}, $|\delta^{\pi\eta\eta'}_\alpha|<1.4^\circ$. 
Additional $\Delta I =5/2$ corrections are hard to calculate, however are 
expected to introduce another uncertainty at this level~\cite{Gardner2}. 

Thus, while a known negative correction of $-(1.7\pm 0.3)^\circ$ from EWP 
contributions should be included in the isospin-extraced value of $\alpha$ in 
$B\to\pi\pi$ and $B\to\rho\rho$, an uncertainty at this 
level remains in $B\to \pi\pi$ from other isospin breaking terms. 
The extraction of $\alpha$ in $B\to \rho\rho$ involves two additional corrections, 
from $\rho$-$\omega$ mixing~\cite{GZI} and from the $\rho$ width~\cite{FLNQ}. 
Both effects can be included in the extraction of $\alpha$ by adequate measurements 
of  $\pi\pi$ invariant mass distributions. 

\subsection{The phase $\gamma$ in $B^+\to DK^+$}

The processes $B^+\to D^{(*)}K^{(*)+}$ and their charge conjugates provide 
a way for determining $\gamma$ with high precision~\cite{GW}. 
The neutral $D$ meson can be either a 
$\bar D^0$ from $\bar b\to \bar cu \bar s$ or a $D^0$ from $\bar b \to \bar uc\bar s$.
Every hadronic state $f$ accessible to $\bar D^0$ decay is also accessible to $D^0$ 
decay. An interference between the two channels leading to a common state $fK^+$ 
involves the weak phase difference $\gamma$. This phase can be determined by 
decay rate measurements for $B^+\to fK^+$ and $B^-\to fK^-$. Effects from 
$D^0$-$\bar D^0$ mixing are neglibible~\cite{GSZ}.

Since the original suggestion of fifteen years ago several variants have been proposed 
studying mainly three classes of states $f$: CP eigenstates 
(e.g. $K^+K^-$)~\cite{GW}, flavor states (e.g. $K^-\pi^+$)~\cite{ADS} and multi-body 
states (e.g. $K_S\pi^+\pi^-$)~\cite{GGSZ}. In $B^+\to DK^+, D\to f$ the three variants involve 
a common ratio of amplitudes, $A(B^+\to D^0 K^+)/A(B^+\to \bar D^0K^+) \equiv re^{i\delta}e^{i\gamma}$, $r\sim 0.1 - 0.2$, differing by their complex ratio, 
$A(D^0\to f)/A(\bar D^0\to f)$. This ratio is exactly $\pm 1$ and about $\lambda^{-2}$
in the first two variants, and depends on the point in a Dalitz plot in the third variant.

The limiting factor of the method is the small value of $r$, for which the current 
upper limit (at 90$\%$ C.L.)~\cite{r-bound} $r < 0.18$ approaches 
estimates~\cite{MGDK}. A nonzero value of $r$ may be measured soon. The 
corresponding ratio of amplitudes in self-tagged $B^0\to DK^{*0}$
is expected to be larger  than in $B^+\to DK^+$. If that is the case also for 
a ratio $r^*$ in $B^+\to DK^{*+}$ then one may be able to observe soon a difference 
between two ratios, $R^*_{\pm}\equiv \b(D_{CP\pm}K^{*+})/\b(D_{\rm flavor}K^{*+})$ 
which grows linearly with $r^*$.
This is a key step towards measuring $\gamma$ using CP eigenstates~\cite{MGDK}.
Recent measurements~\cite{BaDK*}, $R^*_+=1.96 \pm 0.40 \pm 0.11, R^*_-=0.65 \pm 
0.26 \pm 0.08$, implying a difference at $2.6\sigma$, illustrate the need for somewhat
higher statistics.

The variant which involves the largest statistics studies multi-body Cabibbo-allowed
$D$ decays~\cite{GGSZ}. Defining $m^2_{\pm} \equiv (p_{K_S}+p_{\pi^{\pm}})^2$, 
one writes
\bea
& A(B^+ \to (K_S\pi^+\pi^-)_DK^+) =\nonumber\\
&~~~~f(m^2_+,m^2_-) + re^{i(\delta+\gamma)}f(m^2_-,m^2_+),
\eea
replacing $m_+ \leftrightarrow m_-,~\gamma \to -\gamma$ in $B^-$ decay.
The function $f$ is obtained by modeling separately 
measured flavor-tagged $D^0\to K_S\pi^+\pi^-$ as a sum of about twenty 
resonant and nonresonant contributions~\cite{DalitzBe,DalitzBa}. This 
introduces a certain ambiguity~\cite{ambig} and a model-dependent uncertainty
in the analysis. Fitting the $B^{\pm}\to (K_S\pi^+\pi^-)_DK^{\pm}$ rates for a 
given function $f$ to the parameters $r,\delta, \gamma$, one then determines 
the three parameters. 

Averaging~\cite{DalitzBe} $\gamma=(68^{+14}_{-15}\pm 13\pm 11)^\circ$ 
and~\cite{DalitzBa} $\gamma=(67\pm 28\pm 13\pm11)^\circ$ combining $DK, D^*K,$ $DK^*$,
one finds $\gamma= (68\pm 18)^\circ$. The first errors ($^{+14^\circ}_{-15^\circ}$ and 
$\pm 28^\circ$) depend inversely on $r^{(*)}$, showing the importance of fixing $r^{(*)}$,
a key ingredient also in using CP-eigenstates of $D$. The last errors ($\pm 11^\circ$) from 
modeling $f$ may be reduced by studying at CLEO-c $D_{CP\pm}\to K_S\pi^+\pi^-$, 
which determines strong phases in $D$ decays~\cite{GGRdelta}.

\subsection{Consistency between $\alpha,~\beta$ and $\gamma$}

The direct measurements of the weak phases, $\beta=(21.6 \pm 1.3)^\circ,
\alpha = (97 \pm 8)^\circ~({\rm in}~B\to \pi\pi,~\rho\rho$,
$\rho\pi)$ and $\gamma = 
(68 \pm 18)^\circ~({\rm in}~B\to D^{(*)}K^{(*)})$, imply $\alpha + \beta + \gamma = 
(187 \pm 20)^\circ$. 
Since $\alpha$ in $B\to \pi\pi, \rho\rho, \rho\pi$ is defined as $\pi -\beta-\gamma$, 
the question posed by the sum is actually whether the above measurements of $\beta$ and 
$\gamma$, $\beta+\gamma=(90\pm 18)^\circ$, agree with the measurement 
$\beta+\gamma=(83\pm 8)^\circ$ in $B\to \pi\pi,~\rho\rho,~\rho\pi$. 

The sum $\alpha+\beta+\gamma$ measured in this way does not check the unitarity 
of the $3\times 3$ CKM matrix which is violated in models with additional quarks.
The sum is unaffected by New Physics in $B^0$-$\ob$ mixing,
which contributes equally with opposite signs to $\beta$ and $\alpha$,
nor is it affected 
by New Physics in $\Delta I=1/2$ $b\to dq\bar q$ amplitudes which are eliminated 
in the isospin method~\cite{GL}. The sum would be affected by New Physics in 
$\Delta I=3/2$ $b\to dq\bar q$ transitions. This could lead to nonzero CP asymmetries in 
$B^+\to\pi^+\pi^0$ [currently~\cite{HFAG} $A_{CP}(\pi^+\pi^0)=0.01\pm 0.06$] and 
$B^+\to\rho^+\rho^0$, where the Standard Model predicts a vanishing asymmetry 
including EWP contributions~\cite{BFGPY}. 
Other probes of $\Delta I=3/2$ New Physics in $B\to\pi\pi$ are discussed in~\cite{London}.
The sum $\alpha+\beta+\gamma$ could also be affected by
CP violation in $D^0$-$\bar D^0$ mixing, which can be tested directly 
in $D^0$ decays.

 \section{$\sin 2\beta$ IN $b \to s q\bar q$ DECAYS}

In a class of penguin-dominated $B^0$ decays into CP-eigenstates, including
the final states  $f=\eta' K_S,~\phi K_S,~\pi^0 K_S,~f_0K_S,~\rho^0K_S,~\omega K_S,
K^+K^-$ $K_S$, $K_SK_SK_S, K_S\pi^0\pi^0$, decay amplitudes contain two 
terms: a penguin amplitude, $p_f$, involving a dominant 
CKM factor $V^*_{cb}V_{cs}$,  and another term, $c_f$, with a smaller CKM 
factor $V^*_{ub}V_{us}$. The CP asymmetry involves two 
terms, $S_f\sin\Delta mt-C_f\cos\Delta mt$, for which expressions were derived 
sixteen years ago~\cite{PenMG} for a final state of CP eigenvalue $\eta_f$, and for
a small value of $\xi_f\equiv |c_f|/|p_f|$,
\beq
~~~ C_f  \approx 2\xi_f\sin\gamma\sin\delta_f,
 \eeq
 $$
 \Delta S_f \equiv -\eta_f S_f -\sin 2\beta \approx 2\xi_f\cos 2\beta\sin\gamma\cos\delta_f.
\nonumber
$$
For fixed $\beta$, these equations describe a circle,
\beq
C^2_f  + (\Delta S_f/\cos2\beta)^2 = 4\xi^2_f\sin^2\gamma,
\eeq
on which points are parametrized by $\delta_f$, the strong phase difference 
between $c_f$ and $p_f$.

Predictions for $C_f$ and $\Delta S_f$ require knowing the hadronic quantities 
$\xi_f$ and $\delta_f$. A precise knowledge of $\xi_f$ and $\delta_f$ is crucial for 
claiming evidence for physics beyond the Standard Model in the relevant asymmetry 
measurements. This question has been studied using two major approaches, 
flavor SU(3)~\cite{GHLR} and QCD factorization~\cite{BBNS}. A third approach is 
based on final state rescattering~\cite{CCS}. I will describe the first two
methods sketching their predictions.

Flavor SU(3) has been applied in two ways. In one type of study, decay rates and
CP asymmetries have been correlated successfully for a wide variety of charmless 
$B$ decays involving two light pseudoscalars ($P$)~\cite{CGRS}  and 
a pseudoscalar and a vector meson ($V$)~\cite{CGLRS}. This led to 
predictions for the magnitudes and signs of $C_f$ and $\Delta S_f$, which may involve 
uncertainties at a level of $30\%$ from SU(3) breaking. Note that the sign of 
$\Delta S_f$ is predicted to be positive under a very mild assumption, 
$|\delta_f|< \pi/2$, which holds for several final states including $\pi^0K_S$ and $\eta'K_S$~\cite{CGRS}. 

In a more conservative approach, SU(3) has been used
to relate the amplitudes $p_f$ and $c_f$ to  linear combinations of corresponding 
amplitudes in strangeness conserving decays~\cite{SU3bounds}. The resulting 
prediction for a given final state $f$ is that the point $(C_f,\Delta S_f/\cos 2\beta)$ must  
lie within a circle of a given radius. In this approach the signs of $C_f$ and
$\Delta S_f$ are unpredictable. 

QCD factorization has been applied to $B\to PP$ and $B\to VP$ by
expanding decay amplitudes in $1/m_b$ and $\alpha_s$~\cite{BN}.
Since strong phases are suppressed in this expansion, one expects 
$\Delta S_f>0$ in most cases.
Uncertainties include corrections from nonperturbative charming 
penguin contributions (also interpreted as long distance final state interactions),
and $1/m_b$ terms which may be large. This ambitious approach fails, for instance, 
in $B\to K^*\pi$, where predicted branching ratios are consistently lower than the 
data by a factor two to three. 

A sample of predictions, $\Delta S_f=0.10\pm0.05,~0.03\pm0.02,~0.03\pm0.02$,
for $f=\pi^0K_S,~\eta'K_S,~\phi K_S$, respectively, is common to flavor SU(3) and 
QCD factorization. Asymmetry measurements~\cite{HFAG} updated 
by recent studies~\cite{Belle,Baeta'K} are consistent with these predictions 
and with predictions or bounds on other asymmetries. 
Errors must be reduced by at least a factor two before claiming
evidence for New Physics.

\section{NEW PHYSICS IN $A_{CP}(B\to K\pi)$}

An independent test for New Physics in $b\to sq\bar q~(q=u,d)$ has been proposed
recently in terms of a sum rule among four $B\to K\pi$ CP asymmetries~\cite{KpiAsym}.
The sum rule, obeyed by CP rate differences, 
$\Delta_{ij}\equiv \Gamma(B\to K^i\pi^j) - \Gamma(\bar B\to K^{\bar i}\pi^{\bar j})$,
reads
\beq\label{SR}
\Delta_{+-} + \Delta_{0+} = 2(\Delta_{+0}+\Delta_{00}).
\eeq
This relation, reminiscent of a similar sum rule among partial decay rates~\cite{GRLip},
is more precise than relations omitting $\Delta_{0+}$~\cite{GRDel} and 
$\Delta_{00}$~\cite{GRLip}. It is expected to hold in the Standard Model within a 
few percent. A proof of the sum rule will now be sketched discussing briefly its implication.

The first step of the proof is based on isospin considerations, neglecting 
subleading $\Delta I=1$ electroweak penguin contributions to decay amplitudes~\cite{AS}.
A dominant $\Delta I=0$ penguin term with CKM factor $V^*_{tb}V_{ts}$ is 
common to the four $K\pi$ decay amplitudes, up to a factor $1/\s$ in processes involving 
a  $\pi^0$. This common term interferes in CP rate differences with tree amplitudes involving a 
CKM factor $V^*_{ub}V_{us}$. The difference between the left and right-hand sides of 
(\ref{SR}) consist of an interference with a superposition of amplitudes which vanishes by isospin~\cite{isoKpi}.

The remaining terms in (\ref{SR})  consist of subleading electroweak 
penguin amplitudes interfering with tree amplitudes involving $V^*_{ub}V_{us}$. This 
interference vanishes 
in the flavor SU(3) and heavy quark limits. Here one is using a proportionality relation  
between the strangeness changing $\Delta I=1$ EWP operator and the $\Delta I=1$ 
current-current operator in the effective Hamiltonian~\cite{BFGPY,NR}, and a property,
${\rm Arg}(C/T)\sim{\cal O}(\Lambda_{\rm QCD}/m_b,~\alpha_s(m_b))$, of the ratio of color-suppressed and 
color-favored tree amplitudes~\cite{BPRS}. Terms which are both subleading and 
symmetry breaking are estimated to be a few percent of $\Delta_{+-}$ and are negligible.

Using measured asymmetries in $B\to K^+\pi^-$, $K^+\pi^0,K^0\pi^+$, and 
the four $K\pi$ branching ratios~\cite{HFAG}, one 
predicts~\cite{KpiAsym} $A_{CP}(K^0\pi^0)=-0.17\pm 0.06$, to be 
compared with the current value, $A_{CP}(K^0\pi^0)=0.02 \pm 0.13$. Testing 
New Physics requires reducing the experimental error by at least a factor two. 
The sum rule may be violated by an anomalous 
$\Delta I=1$ EWP-like operator. 
 
 \section{CONCLUSIONS}
 
 CP asymmetries measured in a number of $B$ decays, involving
 a variety of interferences, support the hypothesis that the dominant 
 origin of CP violation is a single phase in the CKM matrix. These 
 measurements have reduced the error on $\beta$ to $\pm1^\circ$, and  
the error on $\alpha$ (or $\gamma$) to $\pm 7^\circ$. 
A correction $\Delta^{EWP}_{\alpha}=-1.7^\circ$ must be included. 
 
 Reducing the error in $\alpha$ depends on improved  
 measurements of $A_{CP}(\pi^0\pi^0)$ predicted to be large and positive~\cite{GRDel},
 and on improved upper bounds on 
 $\b(\rho^0\rho^0)$, for which a nonzero value may be measured soon. Isospin breaking 
 effects on $\alpha$ caused by $\rho$-$\omega$ mixing and by the $\rho$ width should be 
 studied in $\pi\pi$ mass distributions in $B\to \rho\rho$. 
 A precision determination of $\gamma$ in 
 $B\to D^{(*)}K^{(*)}$, by combining CP-eigestates, flavor states and multi-body decays 
 of $D^0$,  may be achieved soon.
 
 Time-dependent CP asymmetries  in $b\to s$ decays 
 converge to Standard Model predictions. Testing New Physics in these decays requires 
 precise calculations of small amplitudes and strong phases. A test using the sum rule for 
 all $B\to K\pi$ asymmetries requires reducing the error in $A_{CP}(K^0\pi^0)$ which is 
 experimentally challenging.

\medskip
I thank Tom Browder and Jonathan Rosner for useful comments.
This work was supported in part by the Israel Science Foundation
under Grant  No. 1052/04 
and by the German-Israeli Foundation under Grant No. I-781-55.14/2003. 

\def \ajp#1#2#3{Am.\ J. Phys.\ {\bf#1} (#3) #2}
\def \apny#1#2#3{Ann.\ Phys.\ (N.Y.) {\bf#1}, #2 (#3)}
\def \app#1#2#3{Acta Phys.\ Polonica {\bf#1}, #2 (#3)}
\def \arnps#1#2#3{Ann.\ Rev.\ Nucl.\ Part.\ Sci.\ {\bf#1}, #2 (#3)}
\def \art{and references therein}
\def \cmts#1#2#3{Comments on Nucl.\ Part.\ Phys.\ {\bf#1}, #2 (#3)}
\def \cn{Collaboration}
\def \econf#1#2#3{Electronic Conference Proceedings {\bf#1}, #2 (#3)}
\def \efi{Enrico Fermi Institute Report No.}
\def \epjc#1#2#3{Eur.\ Phys.\ J.\ C {\bf#1} (#3) #2}
\def \ib{{\it ibid.}~}
\def \ibj#1#2#3{~{\bf#1}, #2 (#3)}
\def \ijmpa#1#2#3{Int.\ J.\ Mod.\ Phys.\ A {\bf#1} (#3) #2}
\def \ite{{\it et al.}}
\def \jhep#1#2#3{JHEP {\bf#1} (#3) #2}
\def \jpb#1#2#3{J.\ Phys.\ B {\bf#1}, #2 (#3)}
\def \mpla#1#2#3{Mod.\ Phys.\ Lett.\ A {\bf#1} (#3) #2}
\def \nat#1#2#3{Nature {\bf#1}, #2 (#3)}
\def \nc#1#2#3{Nuovo Cim.\ {\bf#1}, #2 (#3)}
\def \nima#1#2#3{Nucl.\ Instr.\ Meth.\ A {\bf#1}, #2 (#3)}
\def \npb#1#2#3{Nucl.\ Phys.\ B~{\bf#1}  (#3) #2}
\def \npps#1#2#3{Nucl.\ Phys.\ Proc.\ Suppl.\ {\bf#1} (#3) #2}
\def \PDG{Particle Data Group, K. Hagiwara \ite, \prd{66}{010001}{2002}}
\def \pisma#1#2#3#4{Pis'ma Zh.\ Eksp.\ Teor.\ Fiz.\ {\bf#1}, #2 (#3) [JETP
Lett.\ {\bf#1}, #4 (#3)]}
\def \pl#1#2#3{Phys.\ Lett.\ {\bf#1}, #2 (#3)}
\def \pla#1#2#3{Phys.\ Lett.\ A {\bf#1}, #2 (#3)}
\def \plb#1#2#3{Phys.\ Lett.\ B {\bf#1} (#3) #2}
\def \prl#1#2#3{Phys.\ Rev.\ Lett.\ {\bf#1} (#3) #2}
\def \prd#1#2#3{Phys.\ Rev.\ D\ {\bf#1} (#3) #2}
\def \prp#1#2#3{Phys.\ Rep.\ {\bf#1} (#3) #2}
\def \ptp#1#2#3{Prog.\ Theor.\ Phys.\ {\bf#1} (#3) #2}
\def \rmp#1#2#3{Rev.\ Mod.\ Phys.\ {\bf#1} (#3) #2}
\def \rp#1{~~~~~\ldots\ldots{\rm rp~}{#1}~~~~~}
\def \yaf#1#2#3#4{Yad.\ Fiz.\ {\bf#1}, #2 (#3) [Sov.\ J.\ Nucl.\ Phys.\
{\bf #1}, #4 (#3)]}
\def \zhetf#1#2#3#4#5#6{Zh.\ Eksp.\ Teor.\ Fiz.\ {\bf #1}, #2 (#3) [Sov.\
Phys.\ - JETP {\bf #4}, #5 (#6)]}
\def \zp#1#2#3{Zeit.\ Phys.\ {\bf#1} (#3) #2}
\def \zpc#1#2#3{Zeit.\ Phys.\ C {\bf#1} (#3) #2 }
\def \zpd#1#2#3{Zeit.\ Phys.\ D {\bf#1}, #2 (#3)}

\end{document}